\documentclass{article}
\usepackage{algorithm}
\usepackage[noend]{algorithmic}
\usepackage{subcaption} 
\usepackage{hyperref}
\usepackage{cleveref}
\usepackage{listings}
\usepackage{siunitx}
\usepackage{graphicx}
\hypersetup{pageanchor=false}
\renewcommand{\emph}[1]{{\textbf{\textit{#1}}}}
\newcommand{\ignore}[1]{\relax}

\begin{document}

\title{\bf \vspace{-2cm}Modeling the Small-World Phenomenon with Road Networks}
\author{Michael T. Goodrich\thanks{Dept. of Computer Science, University of California, Irvine, \nolinkurl{{goodrich,eozel}@uci.edu}} \and Evrim Ozel\footnotemark[1]}
\date{}
\maketitle

\begin{abstract}
Dating back to
two famous experiments by the social-psychologist, Stanley Milgram,
in the 1960s,
the \emph{small-world phenomenon} is the idea that all people are connected
through a short chain of acquaintances that can be used to route messages.
Many subsequent papers have attempted to model this phenomenon, 
with most concentrating
on the ``short chain'' of acquaintances rather than their ability to
efficiently route messages.
For example, a well-known \emph{preferential attachment} model
by Barab{\'a}si and Albert provides a mathematical explanation of how
a social network can have small diameter---hence, short chains between
participants---but this model doesn't explain how they 
can route messages.
A notable exception is a well-known model by Jon Kleinberg, 
which shows that it is possible for participants in a $n\times n$ grid to route
a message in $O(\log^2 n)$ hops by augmenting the grid with a small number
of long-range random links and using a simple greedy routing strategy.
Although Kleinberg's model is intriguing, 
it does not take into account the road network of the United States used
in the original Milgram experiments and
its $O(\log^2 n)$ number of hops for messages is actually 
quite far from the average of six hops for successful
messages observed by Milgram in his experiments,
which gave rise to the ``six-degrees-of-separation'' expression.
In this paper, we study the small-world
navigability of the U.S.~road network, with the goal of providing
a model that explains how
messages in the original small-world experiments could be routed
along short paths using U.S.~roads.
To this end, we introduce the 
\emph{Neighborhood Preferential Attachment} model, which combines
elements from Kleinberg's model
and the Barab{\'a}si-Albert model, such that long-range links are chosen
according to both the degrees and (road-network) 
distances of vertices
in the network. We empirically evaluate all three models by running
a decentralized routing algorithm, where each vertex only has
knowledge of its own neighbors, and find that our model outperforms
both of these models in terms of the average hop length. Moreover,
our experiments indicate that similar to the Barab{\'a}si-Albert model,
networks generated by our model are scale-free, which could be a
more realistic representation of acquaintanceship links in the
original small-world experiment.
\end{abstract}

\section{Introduction}
The \emph{small-world phenomenon} is the idea
that all people are connected through a short chain of acquaintances that can
be used to route messages.
This phenomenon was popularized 
by the social-psychologist, Stanley Milgram, based on two experiments 
performed in the 1960s~\cite{milgram1967small,travers1969experimental},
where a randomly-chosen group of people 
were given packages to send to someone in Massachusetts.
Each participant was told that they should mail their package only to
its target person if they knew them on a first-name basis; otherwise, they
should mail their package to someone they knew who is more likely
to know the target person.
Remarkably, many packages made it to the target people, with the
median number of hops being 6,
which
gave rise to the expression that everyone is separated by
just ``six degrees of separation''~\cite{guare1990six}.

\begin{figure}[hbt]
\centering
\includegraphics[width=\linewidth]{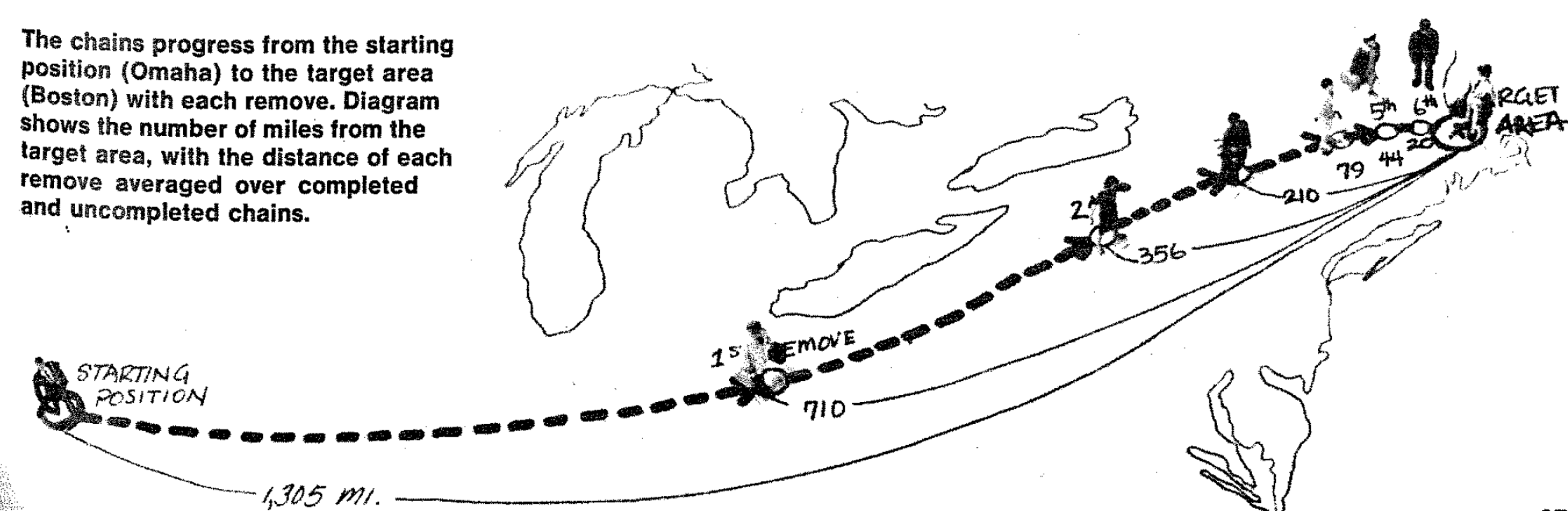}
\caption{\label{fig:hops} Illustration of geographic data
from an original small-world experiment, from~\cite{milgram1967small}.}
\end{figure}

Subsequent to this pioneering research, many papers have
been written on the small-world phenomenon,
e.g., see~\cite{easley2010networks}, with a number of models having
been proposed to explain it.
Nevertheless, based on our review of the literature, the models proposed
so far do not fully explain observations
made by Milgram 
regarding his experiments~\cite{milgram1967small,travers1969experimental}.
For example, Milgram observed that message routing occurred 
in a geographic setting with distances (measured in miles, presumably
in the road network of the United States) roughly halving with each hop;
see Figure~\ref{fig:hops}.

In spite of the geographic nature of the early small-world 
experiments,\footnote{
  The first experiment involved a group of people in Wichita, Kanasas 
  who were asked to send a package to the wife of a divinity student 
  in Cambridge, and the second experiment involved a group of people 
  in Omaha, Nebraska (plus a small number of folks in Boston) who were 
  asked to send a package to a stock broker who worked in Boston
  and lived in Sharon, Mass~\cite{milgram1967small}.}
we are not familiar with any previous work that models the small-world
phenomenon with road networks.
Thus, we are interested in this paper in modeling the small-world phenomenon
with road networks.
For example,
one of the surprising results in the original small-world experiments
was that people were able to find very short paths among acquaintances
with only a limited knowledge of the social network of acquaintances.  
This suggests that a model should
explain how people can find short paths in a social network using a
decentralized greedy algorithm, where individuals, who only have
knowledge of their direct acquaintances, attempt to send a message
towards a target along some path.

\subsection{Related Prior Work}
Arguably, the closest prior
work on a model directed at explaining how small-world (social-network)
greedy routing can work in a geographic setting
is a well-known model by Jon Kleinberg~\cite{DBLP:conf/stoc/Kleinberg00}.
Rather than using a road network, however,
Kleinberg's model is built on a two dimensional
$n\times n$ grid, where each grid point corresponds to a single person,
with two types of 
connections---\emph{local connections} and \emph{long-range connections}. 
The local connections of the network are made by connecting each grid
point to every other grid point within lattice distance $p\geq 1$.
The long-range connections are made by connecting each grid point
to $q \geq 0$ other grid points chosen randomly 
(typically with $q=1$ or $q$ being a small constant), such that the
probability that grid point $u$ is connected to grid point $v$ is
proportional to $[d_h(u, v)]^{-s}$, where $d_h(u, v)$ is the lattice
distance between $u$ and $v$, and $s$ 
is the \emph{clustering exponent} of the network.
Kleinberg showed that in an $n \times n$ grid, 
a decentralized greedy algorithm, where each message holder
forwards its message to an acquaintance that is closest to the
target grid point, is able to achieve an expected path length of
$O(\log^2 n)$ for $p=q=1$ and $s=2$, with a constant 
of at least $88$ in the leading term in his Big-O 
analysis~\cite{DBLP:conf/stoc/Kleinberg00}.

\begin{figure}[hbt]
\centering
\includegraphics[width=\columnwidth, trim=1.5in 0in 2in 0in, clip]{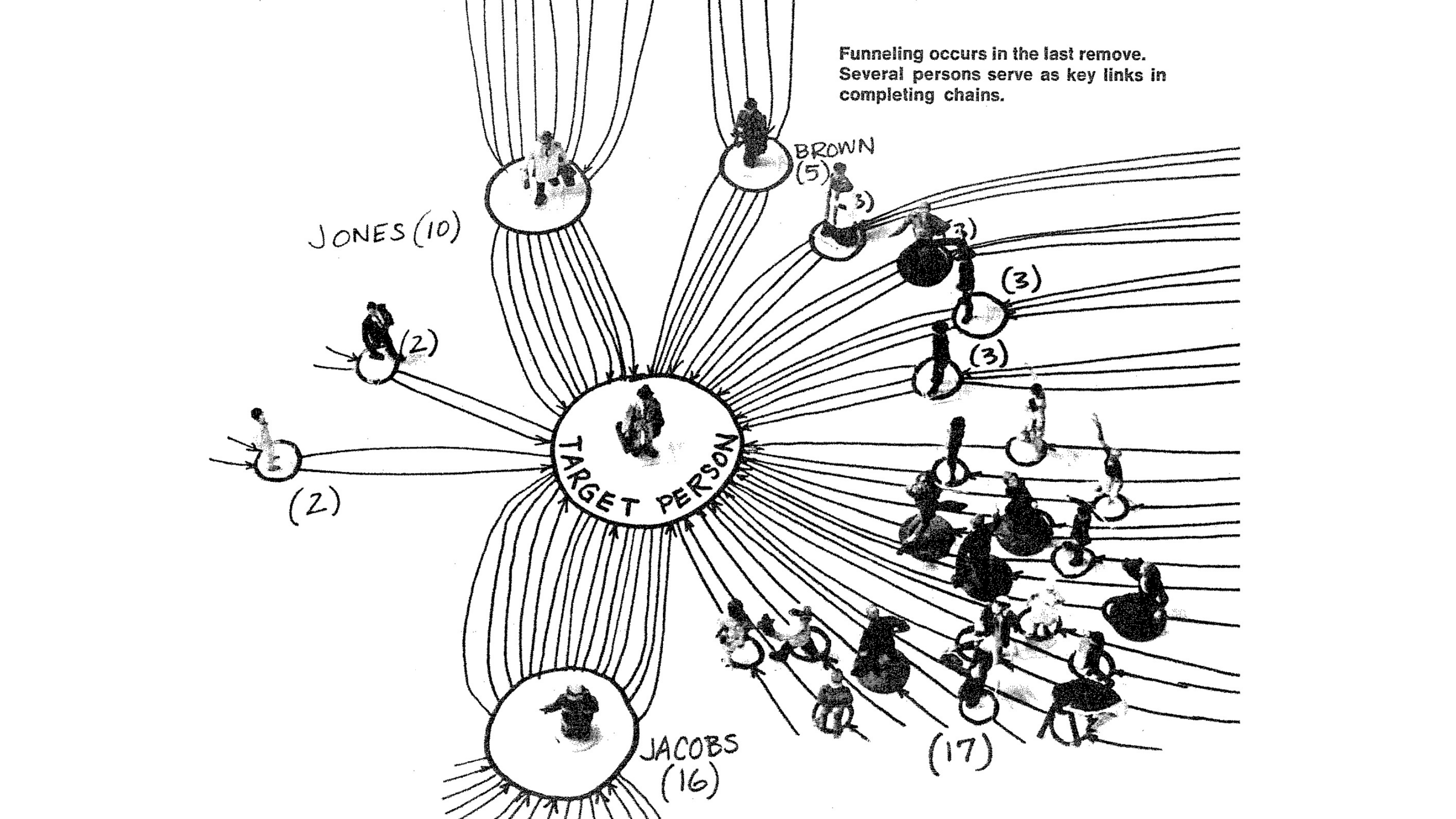}
\caption{\label{fig:people} Final hops for the paths of
delivered packages for people
in an original small-world experiment, from~\cite{milgram1967small}.
Roughly half of the paths were routed through three ``key'' individuals,
Jacobs, Jones, and Brown.}
\end{figure}

When attempting to model the original small-world experiments, however,
there are a number of
drawbacks with the Kleinberg model. First, it requires that the underlying
distances are in the form of a grid, which
is not compatible with how messages were sent in the original small-world
experiments, where messages were sent using the U.S.~road network.
Second, the
upper bound $O(\log^2 n)$, with a hidden constant that is at least $88$,
for the expected number of hops between
vertices does not match the average hop length of six
obtained in the original small-world experiments.
For example, if $n=\,$9,000, then $88\log^2 n$ is approximately 15,000.
Finally, as
we show in \Cref{sec:experiments}, when acquaintanceship links
are viewed as bidirectional, the maximum degree in the resulting
network for the Kleinberg model is quite small. Having a degree
distribution with a heavier tail might be more realistic for a
social network. Moreover, these high-degree vertices might improve
the performance of the model during the routing step. 
Indeed, Milgram noted that in one of his experiments
half of the successfully delivered packages
were routed through three ``key'' 
individuals; see Figure~\ref{fig:people}.

Another well-known social-network model
is the \emph{preferential attachment} model,
which is a random graph model for non-geographic
social networks, such as the World Wide Web. 
This model traces its roots back roughly 100 years, e.g.,
see~\cite{Y25, de_solla_price_general_1976, 10.1093/biomet/42.3-4.425},
and was popularized and formalized by Barab{\'a}si
and Albert~\cite{doi:10.1126/science.286.5439.509}, who also coined the term \emph{scale-free}, which describes networks where the fraction of vertices with degree $d$ follows a power law, $d^{-\alpha}$, where $\alpha>1$.
A graph in the preferential attachment model
is constructed incrementally, starting from a constant-sized ``seed''
graph, adding vertices one-at-a-time,
such that when a vertex, $v$, is added one adds a fixed number, $m$,
of edges incident to $v$, where each other neighbor is chosen with probability
proportional to its degree at that time, e.g.,
see~\cite{batagelj2005efficient}.
This is often called a ``rich-get-richer'' process, and
a rigorous analysis on the degree distribution and
diameter of this model was studied by Bollabas and Riordan~\cite{BR05}.
Further,
Dommers, Hofstad and Hooghiemstra~\cite{Dommers_2010} investigated
the diameters of several variations of the preferential attachment
model, proving that, for each variant, when the power law exponent
exceeds 3, the diameter is $\Omega(\log n)$, and when the power law
exponent is in $(2, 3)$, the diameter is $\Omega(\log\log n)$.

To our knowledge, there does not exist any prior work combining a
preferential attachment model with Kleinberg's model. In terms of
the most relevant prior work, Flaxman, 
Frieze, and Vera~\cite{DBLP:journals/im/FlaxmanFV07} 
introduce a random graph model
that combines preferential attachment graphs with geometric random
graphs, with points created randomly on a unit sphere
one-at-a-time,
such that for each added vertex, $m$ neighbors that are within
a fixed distance, $r$, of that vertex are chosen with probability proportional
to their degrees. Flaxman, Friex, and Vera show that with high
probability the vertex degrees in this model follow a power 
law assuming $r$ is sufficiently large,
and they prove that the diameter of this graph
model is $O(\ln n/r)$ w.h.p., but they do not study its ability to support
efficient greedy routing.
Indeed, when $r\ge\pi/2$, this model is just the preferential attachment
model.

\subsection{Additional Prior Work}
Ever since being popularized by Milgram's experiments and the subsequent work by other researchers on complex networks, the small-world phenomenon has found applications in a wide array of research fields, including rumor spreading, epidemics, electronic circuits, wireless networks, the World Wide Web, network neuroscience, and biological networks. For an overview of the small-world phenomenon and its applications, the reader can refer to~\cite{10.2307/j.ctv36zr5d}.

Incidentally, and not surprisingly,
there has been a significant amount of additional prior work that analyzes the
small-world phenomenon on different types of social network models,
e.g.,  see~\cite{DBLP:journals/pnas/Liben-NowellN0R05,
DBLP:conf/podc/Slivkins05, DBLP:conf/esa/KumarLT06}). 
Liben-Nowell, Novak, Kumar, Raghavan, and Tomkins~\cite{DBLP:journals/pnas/Liben-NowellN0R05} introduce
a geographic social network model, which uses
\emph{rank-based friendships}, where the probability 
of assigning long-range connections
from any person $u$ to person $v$ is inversely proportional to the
number of people in the network who are geographically closer to
$u$ than $v$. The social network is modeled based on a 2D grid
representation of the surface of earth, where each grid point has
a positive population value, and has local connections to its
immediate neighbors on the grid. Each grid point is then connected
to a fifth neighbor based on their rank. Liben-Nowell {\it et al.} 
prove an upper
bound of $O(\log^3 n)$ for the expected hop length of paths formed
by this model, which, of course, is worse than the expected $O(\log^2 n)$
hop lengths in Kleinberg's model.

Kleinberg's model and its extensions have also been studied extensively.
Martel and Nguyen~\cite{DBLP:conf/podc/MartelN04} proved 
the expected
diameter of the resulting graph is $\Omega(\log n)$, but that a greedy
routing strategy cannot find such short paths,
as they show 
that Kleinberg's $O(\log^2)$ analysis for greedy routing is tight.
They extend Kleinberg's model by assuming each vertex has some additional
(unrealistic) knowledge of the network.
For example,
they show that when each node $u$ knows the long-range contacts of
the $\log n$ nodes closest to $u$ in the grid, the expected number
of hops is $O(\log^{3/2} n)$. Fraigniaud, Gavoille and Paul
\cite{DBLP:conf/podc/FraigniaudGP04} provide a similar extension,
and they prove a bound of $O(\log^{1+1/d} n))$ expected hops in the general
$d-$dimensional mesh, and show that this bound is tight for a variety
of greedy algorithms, including those that have global knowledge
of the network.

\subsection{Our Contributions}
In this paper, we study the small-world phenomenon with road networks,
which is motivated by the fact that, as mentioned above,
the network of connections in the original small-world experiments 
were as much geographic as they were 
social~\cite{milgram1967small,travers1969experimental}.
We introduce a new small-world model, which we call the 
\emph{Neighborhood Preferential Attachment} model,
which blends elements from 
the preferential attachment model 
of Barab{\'a}si and Albert~\cite{doi:10.1126/science.286.5439.509}
and Kleinberg's model~\cite{DBLP:conf/stoc/Kleinberg00},
but with underlying distances
defined by a road network rather than a square grid.

In a nutshell,
our model generates a random social network starting from a road network.
We add the vertices to our model one-at-a-time at random from 
the vertices of the underlying road network (whose vertices
stand in as the participants in our social network).
When we add a new vertex, $v$, to our model,
we create a fixed contant number, $m\ge1$, of additional
edges from $v$ to existing vertices, with each other neighbor, $w$, 
chosen with a probability proportional 
to the ratio of the current degree of $w$ (counting just the added edges)
and $d(v,w)^2$, where $d(v,w)$ is the distance from $v$ to $w$ in the
road network.

By using the constant, $m$, as parameter, we guarantee that the average
degree in the network is a constant, which 
matches another observation made by Milgram 
for his experiments~\cite{milgram1967small}.
Interestingly, researchers have observed that an upper bound of
$O(\log n)$ on the expected hop length in Kleinberg's model
can be achieved by having an unrealistic $O(\log n)$ outgoing links for 
\emph{every} vertex instead of a small constant, 
e.g., see~\cite{DBLP:conf/podc/MartelN04}.
Thus, our model tests whether short paths can be found using greedy routing
in a social network with constant average degree, but with a few vertices
having degrees higher than this, as was the
case for the few ``key'' individuals, Jacobs, Jones, and Brown, in an
original small-world
experiment~\cite{milgram1967small}.

One of the main goals
in our design of the Neighborhood Preferential Attachment model
is to introduce a model that brings
the average hop length for greedy routing closer to 
the six degrees-of-separation found in the
original small-world experiments, while keeping the average degree of
the network bounded by a constant.
To test this, we experimentally evaluate instances of our model using
road networks for various U.S.~states.
We empirically compare the performance of greedy routing in our model to 
the performance for
a variant of Kleinberg's model,
where links are chosen with probability proportional to the inverse
squared road-network distances of vertices (rather than a grid), 
as well as with the well-known Barab{\'a}si-Albert 
preferential-attachment model.
Interestingly, our experiments
show that the Neighborhood Preferential
Attachment model outperforms both the 
Barab{\'a}si-Albert preferential-attachment model and the
road-network variant of Kleinberg's model.
Moreover, our experimental results show that our model has a
scale-free degree distribution, 
which is arguably a better representation of real-world
social networks than Kleinberg's model while also being geographic, unlike
the preferential-attachment model of
Barab{\'a}si and Albert.

\section{Preliminaries}
\label{sec:prelim}
We view road networks as undirected, weighted, and connected graphs,
where each vertex corresponds to a road junction or terminus, and
each edge corresponds to road segments that connect two vertices.
In our social network model, each junction or terminus in a road
network represents a single person, and each road segment represents
a social connection between two people, which we consider to be the
local connections of the network. Intuitively, our social network model can be seen as a mapping of each person in the population to the road network vertex that is geographically closest to their address. Likewise, an edge $(u, v)$ in the road network represents the existence of social connections between people who were mapped to vertices $u$ and $v$. This is admittedly an approximation for a population distribution, but we
feel it is reasonable for most geographic regions, since population
density correlates with road-network density,
e.g., see~\cite{doi:10.1080/13658816.2012.663918, BF09, RePEc:eee:trapol:v:45:y:2016:i:c:p:209-217}.
Certainly, it is is more realistic than modeling population density
using a uniform $n\times n$ grid, as in Kleinberg's 
model~~\cite{DBLP:conf/stoc/Kleinberg00}. 

The distance between two vertices $u,v\in V$ is denoted 
as $d(u, v)$ and is the total weight of the shortest path between $u$ and
$v$ in the underlying road network. 
The hop distance between two vertices is denoted as $d_h(u, v)$ 
and is the minimum number of hops required to reach $v$ from
$u$, without considering edge weights and including both road-network
edges and additional edges added during model formation.
In all of the social network models we mention in this paper,
we assume all edges are undirected for the sake of distance computations, 
which reflects the notion that friendships are bidirectional.

We define $\deg_G(v)$ to be the degree of~$v$ in a graph, 
$G=(V,E)$, that is, the number of~$v$'s adjacent vertices in $G$.
If $G$ is understood from context, then we may drop the subscript.

\section{The Road-Network Kleinberg Model} \label{sec:KL-model} 
In this section, we introduce a variant of Kleinberg's small-world
model 
adapted so that it works with weighted road networks rather 
than $n\times n$ grids.
We denote this model throughout this paper as the \emph{KL model}.
Interestingly, as we show in our empirical analysis, although this model is 
not as effective for performing greedy routing as our Neighborhood Preferential
Attachment model, it nevertheless is much more efficient in practice 
than the theoretical analysis 
of Kleinberg~\cite{DBLP:conf/stoc/Kleinberg00}
that is based on using $n\times n$ grids would predict.

As mentioned above,
Kleinberg's network model begins by defining a set of
vertices as the lattice points in an $n\times n$ grid, i.e.,
$\{(i, j) \mid i \in \{1, 2, \ldots, n\}, j \in \{1, 2, \ldots n\}\}$, 
so that the distance between any two vertices $u=(i,j)$ and $v=(k,l)$
is the Manhattan distance, $d(u,v) = |k-i| + |l-j|$. 
Each vertex, $u$,
has an edge to every vertex within distance $p \geq 1$,
called the \emph{local contacts} (typically, we just take $p=1$, so
these are just grid-neighbor connections), 
and each vertex has edges
to $m\geq 1$ other vertices selected at random, called the
\emph{long-range contacts}, such that the probability that there
exists an edge from $u$ to $v$ is $d(u, v)^{-s}/z$, where
$s\geq 0$ is called the \emph{clustering exponent} 
and $z$ is a
normalizing factor that ensures we have a probability distribution. 
Then, a decentralized \emph{greedy algorithm} is used to route
messages between a source and target vertex as follows: at each
step, the current message holder forwards its message to a contact
that has the smallest Manhattan distance to the target vertex.

We now adapt this model to the
KL model that works on weighted road networks.
We start with the set of vertices and edges of a road network, 
where each edge corresponds to a \emph{local} connection. Then, for each
vertex, $u$, we add $m \geq 1$ long-range edges randomly, where the
probability that there exists a long-range connection between
$u$ and a vertex, $v$, is 
$d(u, v)^{-s}/z$, where 
$d(u,v)$ is the road-network distance between $u$ and $v$ 
(in miles or kilometers),
$s \geq 0$ is the clustering exponent,
and $z$ is a
normalizing factor that ensures we have a probability distribution. 
See Algorithm~\ref{alg:kl}, noting that we call it for a road network,
$G=(V,E)$, and parameter, $m\ge 1$, for the number of long-range connections
to add for each vertex.

\begin{algorithm}[hbt]
\begin{algorithmic}[1]
\label{alg:reconstruct}
  \STATE{$E' \gets \emptyset$}
  \FOR{each $v \in V$}{
    \STATE{$P \gets \{1/d(v, u)^s \mid u \in V, u\neq v\}$}
    \STATE{$z_v\gets \sum_{p\in P} p$}
    \STATE{Normalize $P$ by dividing each $p\in P$ by $z_v$}
   \STATE{$S \gets$ sample $m$ vertices according to their probabilities in $P$}
    \STATE{$E' \gets E' \cup \{(v, w) \mid w \in S\}$}
  }
  \ENDFOR
  \RETURN $G=(V,E\cup E')$
\end{algorithmic}
\caption{\textsc{Construct-KL($V, E, s, m$)} \label{alg:kl}}
\end{algorithm}

For his original model (on an $n\times n$ grid),
Kleinberg~\cite{DBLP:conf/stoc/Kleinberg00} showed that the optimal
value for the clustering exponent $s$ is 2, for which the decentralized
greedy
routing algorithm is able to find paths of length $O(\log^2 n)$ in
expectation, and that for any other value of $s\neq 2$, the greedy algorithm
would only be able to find a path with length that is lower bounded
by a polynomial in $|V|$. 
Following Kleinberg,
we usually select $s=2$ for the weighted road-network 
variant, KL, of this model, as well as for the Neighborhood Preferential Attachment model, and we include some experiments that show the
effect of varying this parameter for the latter model on different road networks.

In the routing algorithm for the KL model, 
we use a weighted version of the decentralized
greedy algorithm, such that at each step, the current message holder
forwards its message to a directly adjacent 
contact in the social network that has the smallest
road-network distance to the target vertex (which could have easily been estimated
in the 1960s using a road atlas of the United States and which can be determined
in modern times from any navigation app, such as Google Maps, OpenStreetMap, 
Apple Maps, or Waze).
We denote this greedy algorithm as
\lstinline{Weighted-Decentralized-Routing}. 
\section{A Road-Network Preferential Attachment Model}
\label{sec:BA-model}
In this section,
we give a brief description of the preferential-attachment model;
see, e.g., 
\cite{mitzenmacher2004brief,doi:10.1126/science.286.5439.509,
Dommers_2010,BR05}.
This model is defined by an algorithm to generate random graphs
whose degree distribution follows a power law. The
algorithm is based on a preferential attachment mechanism, where
vertices with larger degrees are more likely to receive new links.

The algorithm for building an instance of the preferential-attachment model
starts with a set, $V$, of $n$ vertices, and
an initial clique of $m+1$ vertices 
from~$V$.\footnote{There are other variations
  for the starting ``seed'' graph, but the results in the limit
  are similar~\cite{mitzenmacher2004brief}.}
It then selects the remaining vertices from $V$ in random order, with each
vertex, $v$, getting connected to $m$ existing vertices,
where the probability that $v$ connects to
vertex $u$ is proportional to $u$'s degree
at the time $v$ is added.
In the case of $m\geq 2$, edges for a particular 
vertex are added through independent trials, i.e.,
previous edges do not affect the degree 
counts when choosing later edges for the same vertex.
The algorithm stops when it has constructed a graph with $n$ vertices.
Note that the number of added edges is exactly~$nm$.
See Algorithm~\ref{alg:ba}.

\begin{algorithm}[hbt]
\begin{algorithmic}[1]
  \STATE{Select subset $M \subseteq V$ of size $m+1$ by sampling vertices u.a.r.}
    \STATE{$E' \gets \{(u,v)\mid u, v \in M, u \neq v\}$}
  \FOR{each $v \in V \setminus M$ in random order}{
    \STATE{$P \gets \{\deg_{G'}(u) \mid u \in V, u\neq v\}$, 
           where $G'=(V,E')$}
    \STATE{$z_v\gets \sum_{p\in P} p$}
    \STATE{Normalize $P$ by dividing each $p\in P$ by $z_v$}
    \STATE{$S \gets$ sample $m$ vertices according to their probabilities in $P$}
    \STATE{$E' \gets E'\cup \{(v, w) \mid w \in S\}$}
  }
  \ENDFOR
  \RETURN $G=(V, E\cup E')$
\end{algorithmic}
\caption{\textsc{Construct-BA($V, E, m$)}\label{alg:ba}}
\end{algorithm}

Although the preferential attachment model is defined as a non-geographic
model, if the vertices in the model have geographic coordinates, such as 
determined in
a road network, we can nevertheless apply the same distributed greedy routing
algorithm as for the KL model. 
Specifically, if we take the set of candidate vertices in the preferential 
attachment model to be vertices in a road network and we union the edges of 
the final preferential attachment model with the edges of the road network
for the corresponding vertices (as shown in Algorithm~\ref{alg:ba}), 
then we can construct an instance
of a preferential-attachment graph embedded in a road network.
This allows each participant to forward their message to a direct contact
(including both added edges and road-network edges)
that is closest to the target (using road-network distance).
Indeed, for our experiments, this is what we
refer to as the \emph{BA model}. 

\section{The Neighborhood Preferential Attachment Model}
We now introduce our
\emph{Neighborhood Preferential Attachment} (NPA) model. 
We start with the
same set of local connections as for 
the road-network Kleinberg model, KL,
except now we distribute long-range connections
according to a combination of 
vertex degrees and road-network distances between vertices.
Thus, our model combines elements of the KL and BA models. 
Surprisingly, as we show below, rather than achieving
a performance somewhere between the KL and BA models,
our NPA model outperforms both the KL model and BA model.

\begin{figure}[hbt]
\centering
\includegraphics[width=\columnwidth, trim = 1.0in 0.5in 0.5in 0.5in, clip]{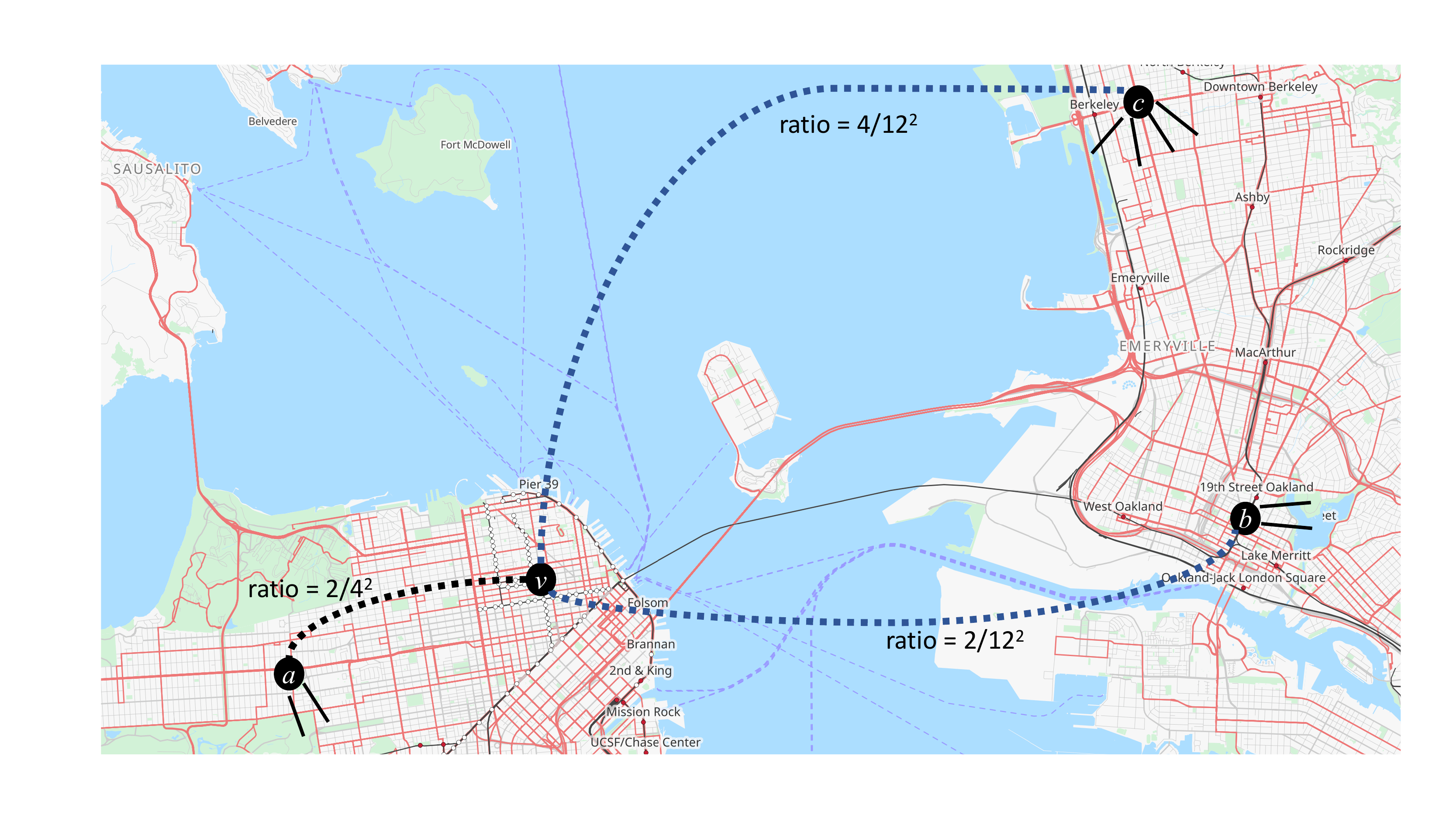}
\caption{\label{fig:npa} How edges are chosen in the Neighborhood 
Preferential Attachment model, illustrated with the road network of 
San Francisco, Berkeley, and Oakland. When vertex $v$ is added,
the ratio for the probability for $a$ is $2/16(=0.125)$,
the ratio for the probability for $b$ is $2/144(=0.014)$,
the ratio for the probability for $c$ is $4/144(=0.028)$.
Thus, even though $b$ and $c$ are the same distance from $v$,
$c$ is twice as likely as $b$ to be chosen, and
$a$ is 4.5 times more likely to be chosen than $c$, because
$c$'s degree of 4 is twice that of $b$ or $a$, but
$a$'s squared distance is 9 times smaller than that of $b$ and $c$.
(Background image is from OpenStreetMap and is 
licensed under the Open Data Commons Open Database 
License (ODbL) by the OpenStreetMap Foundation (OSMF).)
}
\end{figure}

To generate the network of long-range connections, we consider
the vertices in random order, adding
new (long-range) edges, based on degrees, distances,
and an input parameter, $m\ge 1$. 
Let $G=(V,E)$ be a road network of $n$ vertices.
We begin by selecting a subset, $M\subseteq V$, of $m+1$ vertices from $G$
and we add all possible edges between
them, so that every initial vertex has 
an initial degree equal to $m$.
That is, we start by forming a clique of size $m+1$ of randomly
chosen vertices from $V$.
We then repeatedly randomly
consider the remaining vertices from $V$, until we have considered
all the vertices from $V$.
When we process a vertex, $v$, we 
connect $v$ to $m$ other vertices,
where the probability that there is an edge between a new vertex
$v$ and another vertex $u$ is proportional to 
the ratio
$\frac{\deg(u)}{d(v, u)^s}$, 
normalized by normalizing factor, 
\[
z_v = \sum_{w \neq v}\frac{\deg(w)}{d(v, w)^s},
\]
for $v$, such that $\deg(v)$ is the degree of vertex $v$ considering only
added edges and $d(v,u)$ is road-network distance.
Typically, we choose $s=2$.
When $m\geq 2$, 
edges for a particular vertex are added through independent trials.
See Algorithm~\ref{alg:npa} and Figure~\ref{fig:npa}.

\begin{algorithm}[hbt]
\begin{algorithmic}[1]
  \STATE{Select subset $M \subseteq V$ of size $m+1$ by sampling vertices u.a.r.}
    \STATE{$E' \gets \{(u,v)\mid u, v \in M, u \neq v\}$}
  \FOR{each $v \in V\setminus M$ in random order}{
    \STATE{$P \gets \{\deg_{G'}(u)/d(v, u)^s \mid u \in V, u\neq v\}$, where $G'=(V,E')\!\!\!$}
    \STATE{$z_v\gets \sum_{p\in P} p$}
    \STATE{Normalize $P$ by dividing each $p\in P$ by $z_v$}
    \STATE{$S \gets$ sample $m$ vertices according to their probabilities in $P$}
    \STATE{$E' \gets E' \cup \{(v, w) \mid w \in S\}$}
  }
  \ENDFOR
  \RETURN $G=(V,E\cup E')$
\end{algorithmic}
\caption{Construct-NPA($V, E, s, m$)\label{alg:npa}}
\end{algorithm}

Once the model-construction is finished, 
we add the local road-network connections back in. 
Since we add $m$ edges for each vertex in the network, and since
road networks themselves have a constant maximum degree, the average
degree for our network model is a constant when $m$ is a constant.
We refer to this as the \emph{NPA model}.
For the routing phase, we run the same decentralized greedy routing 
algorithm for the NPA model as for the KL and BA models.

\begin{figure*}[!hbt]
\centering
\includegraphics[width=1.07\linewidth, trim=0.5in 0in 0in 0in, clip]{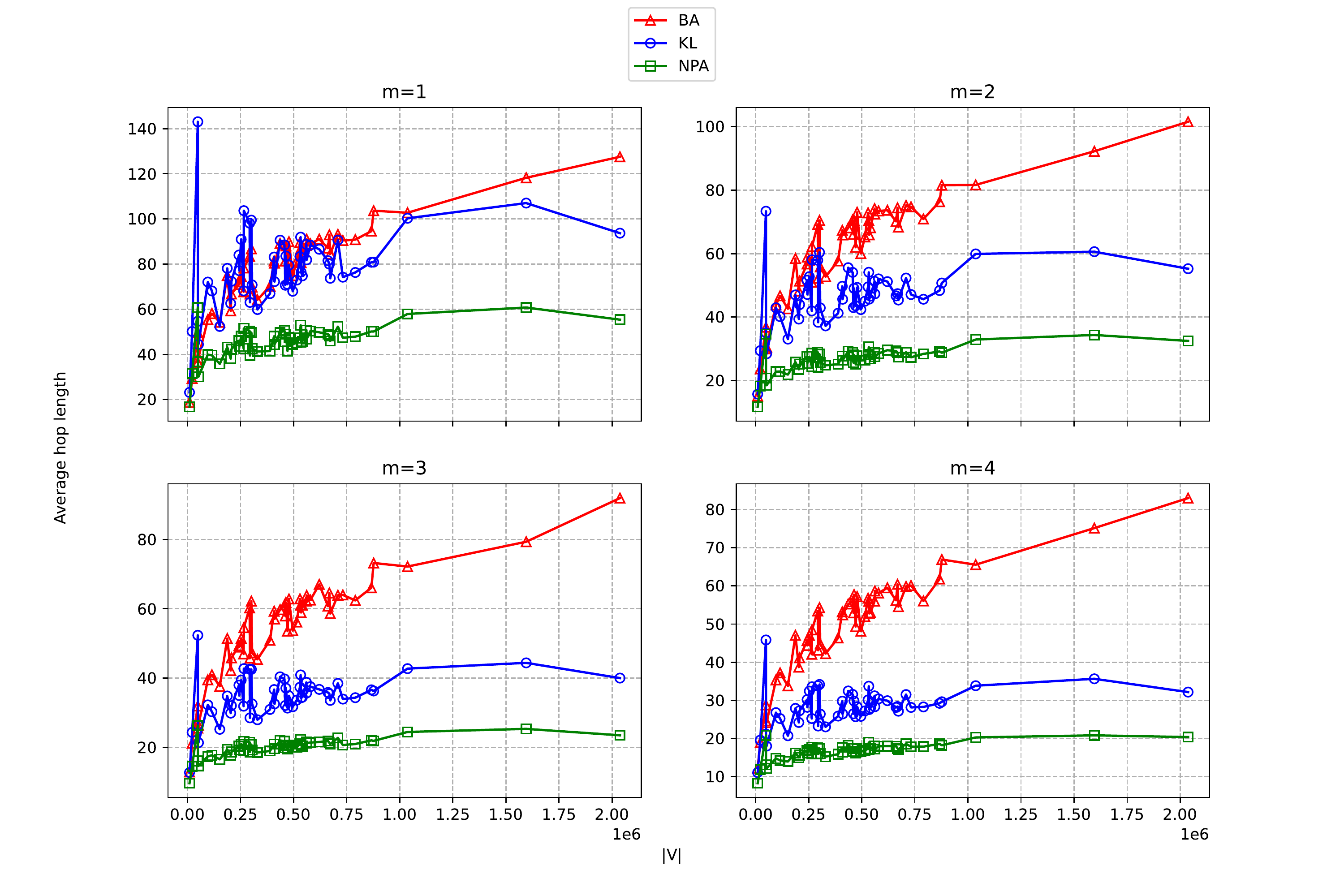}
\caption{Average hop lengths over 1000 runs of \lstinline{Weighted-Decentralized-Routing} for 50 U.S. states and Washington, DC.}
\label{fig:avg-hop-length}
\end{figure*}

\section{Experimental Analysis}
\label{sec:experiments}

Intuitively, the BA model tries to capture how 
popularity is often distributed according to a power law, with
the ``rich getting richer'' as more people are added to a group, but it 
completely ignores geography in forming friendship connections.
That is, in the BA model, if there is a popular person, $u$, in New York and
an equally popular person, $w$, in Los Angeles, a newly-added 
person, $v$, in San Diego is just as likely to form a 
long-range connection to $u$ as to $w$.

The KL model, on the other hand, tries to capture how friendship 
is correlated with geographic distance, but it completely ignores popularity.
That is, in the KL model, if there is a popular person, $u$, in Hollywood and
an unpopular person, $w$, who is also in Hollywood, a newly-added 
person, $v$, in San Diego is just as likely to form a long-range 
connection to $u$ as to $w$.

In contrast to both of these extremes, as illustrated above
in Figure~\ref{fig:npa}, our NPA model tries to capture how friendship 
is correlated with both popularity and geographic distance.
That is, in the NPA model, if there is a popular person, $u$, in New York and
an equally popular person, $w$, in Los Angeles, a newly-added 
person, $v$, in San Diego is more likely to form a long-range 
connection to $w$ than to $u$.
Furthermore, if there is a popular person, $u$, in Hollywood and
an unpopular person, $w$, who is also in Hollywood, a newly-added 
person, $v$, in San Diego is more likely to form a long-range 
connection to $u$ than to $w$.

Intuition aside, however, we are interested 
in this paper in determining
how effective the BA, KL, and NPA models are at greedy routing.
For example,
which of these models is the best at greedy routing and 
can any of them achieve the six-degrees-of-separation phenomenon 
shown in the original small-world 
experiments~\cite{milgram1967small,travers1969experimental}?

\subsection{Experimental Framework}
To answer the above question,
we implemented the BA, KL and NPA models in C++ (using an open-source routing library \cite{DBLP:journals/jea/DibbeltSW16} to find shortest paths), randomly sampled
1000 source/target pairs, then ran
\lstinline{Weighted-Decentralized-Routing} on each pair and measured
the average hop length. The datasets we used are road networks for
50 U.S. states and Washington, D.C., obtained from the formatted
TIGER/Line dataset available from the 9th DIMACS Implementation
Challenge
website.\footnote{\url{http://www.diag.uniroma1.it/~challenge9/data/tiger/}} For each road network, only the largest connected component was considered.
The sizes of the road networks we used range from {9,522} to {2,037,156}
vertices. As a preprocessing step, we normalized edge weights so
that the smallest edge weight is 1.

\subsection{Hop Counts with Few Long-Range Links}
The first set of experiments that we performed was to test the effectiveness
of each of the three models on each road-network data set assuming
that we add only a small number of long-range links.
In particular, we tested each model for the cases when $m=1,2,3,4$.
We show the results of these experiments in \Cref{fig:avg-hop-length},
which show that the NPA model outperforms both the 
KL and BA models for each of these small values for $m$.
For example, even for $m=1$, the number of hops for the NPA model tends
to be half the numbers for the BA and KL models. 
Once $m\ge 2$, the KL model shows improved performance over the BA model,
with the KL model achieving degrees-of-separation values that are
roughly half those for the BA model.
Nevertheless, for $m\ge 2$, the NPA model still beats the KL model, with
hop-counts that are between a third and a half better than the KL model.
Further, as would be expected,
all the models tend to do better as we increase the value of $m$.
For example, when $m=1$, the NPA model achieves 
a degrees-of-separation value of between 40 and 60, 
whereas when we increase $m$ to just $4$, 
the NPA model achieves 
a degrees-of-separation value of between 10 and 20.
Admittedly, this still isn't 6, but it is getting closer, and it
shows what can be achieved with just a few added long-range links.

\subsection{Dropouts}
\begin{figure}[!hbt]
\centering
\includegraphics[width=0.95\columnwidth]{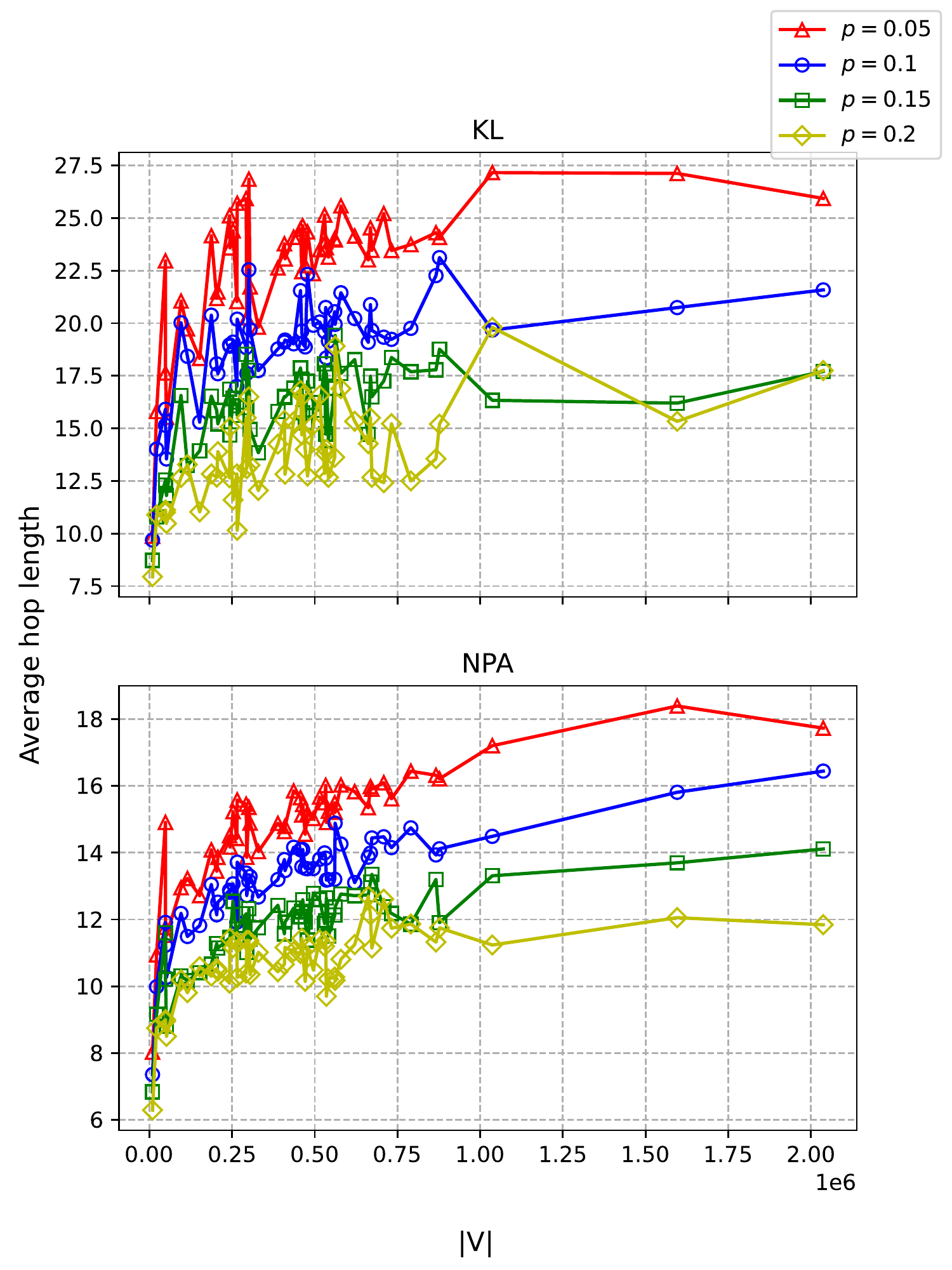}
\caption{Effect of varying the probability $p$ of dropping the message at each step during \lstinline{Weighted-Decentralized-Routing} for the KL and NPA models, with $m=4$.}
\label{fig:drop}
\end{figure}

There is another aspect of the original small-world experiments, which
(like most prior research on the small-world phenomenon)
we have heretofore ignored.
Namely, as participants 
perform greedy routing
in the real world 
there is a probably that someone will simply drop out of the experiment
and not forward the package to anyone.
For example, in one of the original small-world
experiments~\cite{travers1969experimental},
Travers and Milgram observed a dropout probability
of roughly $p=0.2$ at each step in a routing operation.
That is,
in the original small-world experiment, it was observed that some
amount of messages never ended up reaching the target person, e.g.,
due to recipients refusing to participate or not having anyone to
forward the message to.
The longer a source-to-target path gets,
the more likely it is that at least one person will drop the message,
so we expect that the average path length would decrease as the
probability of dropping messages increases. 
To see whether this
could have contributed to the small average hop length observed in
the original small-world experiment, we ran a variant of
\lstinline{Weighted-Decentralized-Routing} on the KL and NPA models,
such that each message holder has a fixed probability $p$ of dropping
the message. Our results can be seen in \Cref{fig:drop}, for $m=4$. 
As expected, these experiments show that the average hop
counts for successful paths decrease as we increase the dropout probability,
$p$, but we still are not quite achieving six degrees of separation for 
these values.

\subsection{Six Degrees of Separation}
\begin{figure}[!hbt]
\centering
\includegraphics[width=\columnwidth]{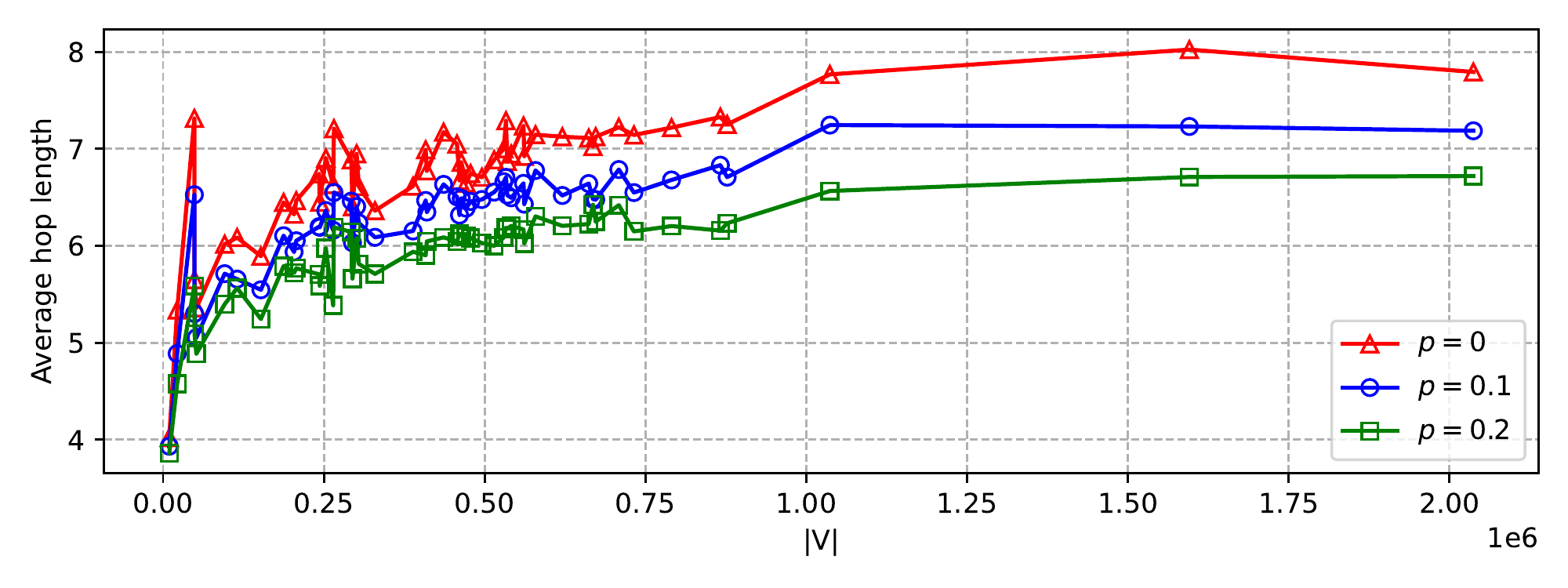}
\caption{Average hop length of the NPA model with $m=30$ for different dropout probabilities.}
\label{fig:30links}
\end{figure}

We can, in fact, achieve 
six degrees of separation in the NPA model, just by slightly increasing
the value of $m$.
In particular, we provide
experimental results in \Cref{fig:30links} for the NPA model with
$m=30$ with different dropout probabilities. 
As this result shows, even with $p=0$ (no dropouts), we can achieve
7 degrees of separation for modestly sized road networks (and 8 degrees
of separation for the three largest road networks).
With $p=0.2$, for the majority of road networks, we get average hop counts that match
the findings in the original small-world experiments, where the
average hop length was found to be 6. For the largest road networks, we get average hop counts that are between 6 and 7.

Intuitively, setting $m=30$ is equivalent to assuming that people
participating in a small-world experiment would consult their address books
when deciding who to send a package to next
and that the average number of entries in each address book is 30, which we
feel is a reasonable assumption.

\section{Diving Deeper}
We are actually interested in more than just showing that the NPA model
can achive six degrees of separation and thereby match the performance of the
original small-world experiments.
In this section, we take a deeper dive into the models we introduce in
this paper, with an eye towards trying to better understand what is going
on during the greedy routing done in each model.

\begin{figure*}[!htb]
\centering
\includegraphics[width=0.95\linewidth]{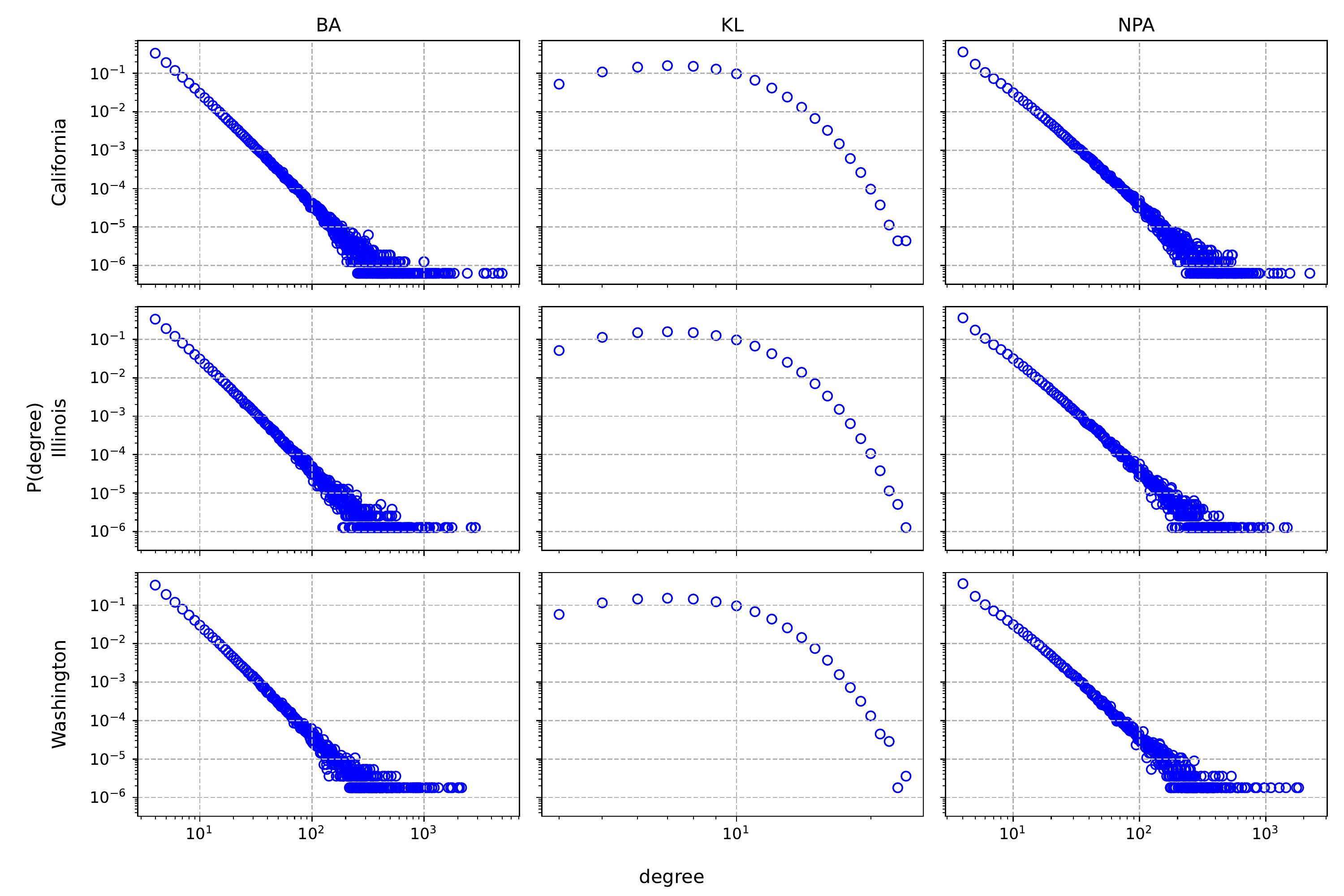}
\caption{Degree distributions of the three main models with $m=4$ on road networks of different sizes.}
\label{fig:dd}
\end{figure*}

\subsection{Degree Distributions}

Comparing the degree distributions
of the three models, which are shown in Figure~\ref{fig:dd},
we see that the KL model has a light-tailed
distribution, whereas our model seems to be scale-free, similar to
the BA model.
These results indicate that the NPA model, similar to the KL model,
is able to utilize local clustering when finding long-range contacts,
while still having the scale-free property. 

\subsection{How Distances to the Target Decrease}
As shown above,
we observe that the NPA model outperforms both of the KL and BA
models in terms of the average hop length. We also see that the KL
model performs significantly better than Kleinberg's theoretical
upper bound~\cite{DBLP:conf/stoc/Kleinberg00} on the grid, which
was $c\log^2 n$ for $c > 88$. 
Still, Kleinberg's theoretical analysis was based on an 
interesting proof technique that was inspired
from Milgram's figure showing how distances to the target tend to halve
with each hop, as shown above in Figure~\ref{fig:hops}.
At a high level, Kleinberg's proof for his $O(\log^2 n)$
bound is based on finding that the probability that the distance from
the current vertex to the target is halved at any step is 
$\Theta(1/\log n)$; hence, this is a constant after $\Theta(\log n)$ hops,
and we can reach the target by repeating this argument $O(\log n)$ times. 

\begin{figure}[!hbt]
\centering
\includegraphics[width=\columnwidth]{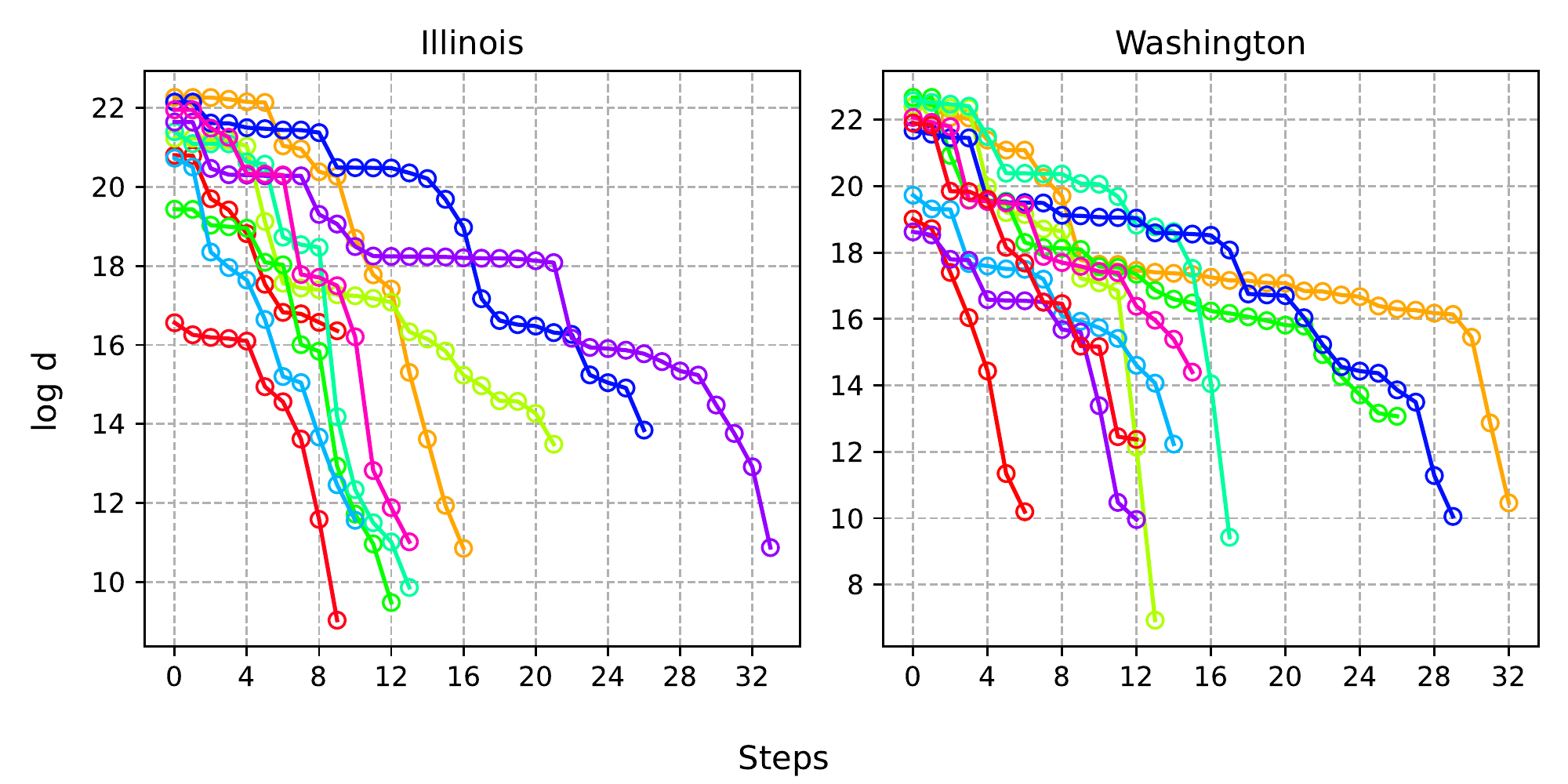}
\caption{Remaining distance to target, denoted as $d$, during 10 runs of \lstinline{Weighted-Decentralized-Routing} on two road networks, with $m=4$. Each line corresponds to a separate run of \lstinline{Weighted-Decentralized-Routing}, with the markers on each line corresponding to the remaining distance at a particular step. The last data point for each run corresponds to the penultimate step, i.e. when the message holder is one hop away from the target.}
\label{fig:halving}
\end{figure}

We provide
experimental results in \Cref{fig:halving} showing how the remaining
distance to the target changes for the NPA model over multiple runs
of \lstinline{Weighted-Decentralized-Routing}. We see that for most
runs, the distance typically gets halved every few steps, as Milgram
observed.

\subsection{Varying the Clustering Coefficient}
\begin{figure}[!hbt]
\vspace*{-10pt}
\includegraphics[width=0.95\columnwidth]{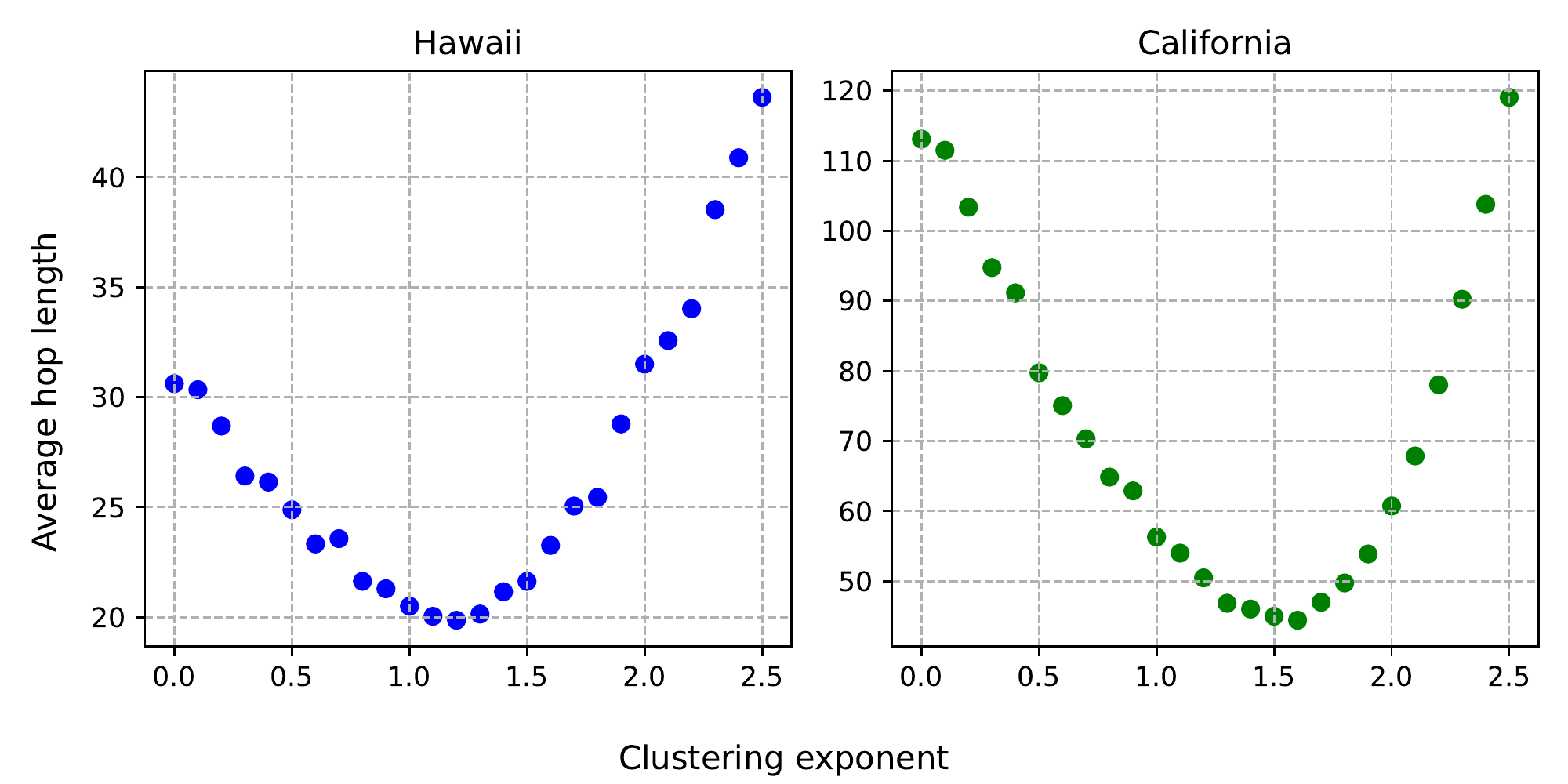}
\vspace*{-10pt}
\caption{Effect of varying the clustering coefficient on the average hop length in the NPA model for the road networks of Hawaii ($|V| = \num{21774}$) and California ($|V| = \num{1595577}$), for $m=1$.}
\label{fig:exp}
\vspace*{-10pt}
\end{figure}

In \Cref{fig:exp}, we see how varying the clustering coefficient affects the average hop length in the NPA model for the HI and CA road networks. Though $s = 2$ is not the best-performing clustering exponent for either road network in our experiments, the results indicate that the best-performing clustering exponent seems to move towards 2 when the input size gets larger, which suggests that the asymptotically optimal clustering exponent could still be 2. A similar effect could be observed in Kleinberg's original model as well, since the lower bounds that are proved for $s\neq 2$ are $\Omega(n^{(2-s)/3})$ for $s<2$ and $\Omega(n^{(s-2)/(s-1)})$ for $s>2$, both of which require input sizes that are orders of magnitude larger than real-world road networks to be able to experimentally observe the optimality of $s=2$.

\subsection{Capping the Maximum Degree}
\begin{figure}[!hbt]
\includegraphics[width=1.02\columnwidth]{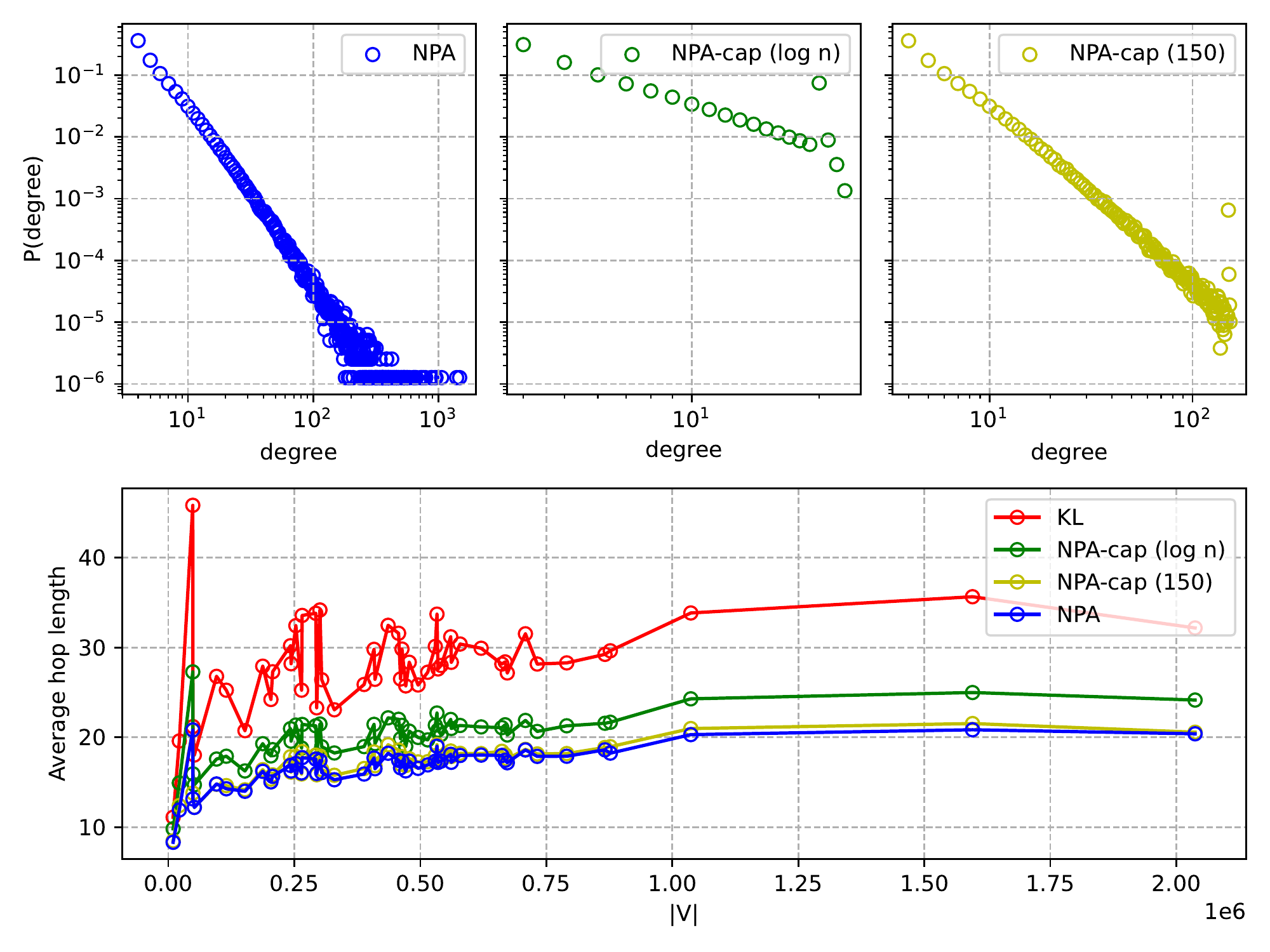}
\caption{Comparing the average hop lengths of the NPA, KL, and the NPA-cap models, and the degree distribution of the NPA and NPA-cap models for 
Illinois, with $m=4$.}
\label{fig:cap}
\end{figure}

We considered another variation of the NPA model, where we cap the
maximum degree such that only vertices of degree less than $c$ are
considered when choosing long-range contacts. We call
this the \emph{NPA-cap} model. 
We choose $c = \log n$ and $c=150$ as possible maximum degree caps.
Intuitively, the cap on the maximum degree is like 
a cap on the size of someone's
address book during a small-world experiment.
We provide experimental results comparing the models KL, NPA, and NPA-cap (for $c=\log n$ and $c=150$), with $m=4$, in \Cref{fig:cap}. 

\begin{figure}[!hbt]
\includegraphics[width=1.02\columnwidth]{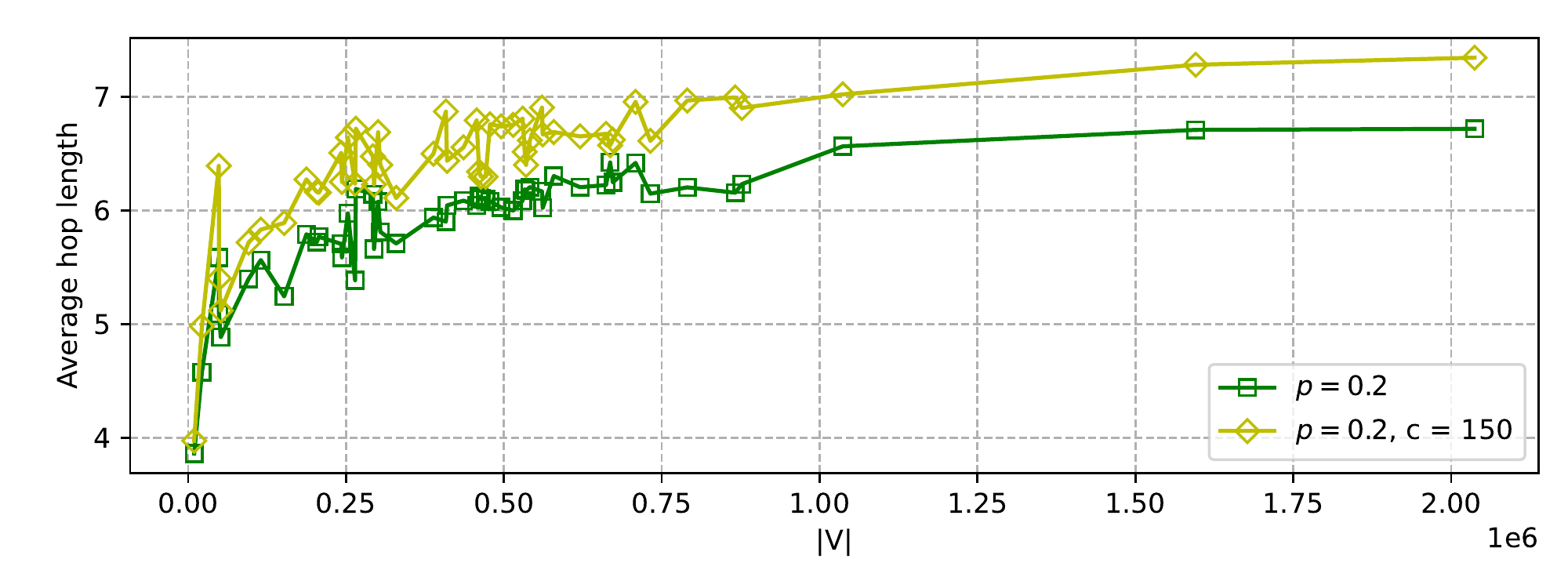}
\caption{Comparing the average hop lengths of the NPA and NPA-cap (150) models with a dropout probability of 0.2 and $m=30$.}
\label{fig:30link-cap}
\end{figure}

In \Cref{fig:30link-cap}, we compare the models NPA and NPA-cap (for $c=150$), when there is a dropout probability of $p=0.2$, with $m=30$.

\subsection{Routing Across Multiple States}
\begin{figure}[!hbt]
\centering
\includegraphics[width=0.75\columnwidth]{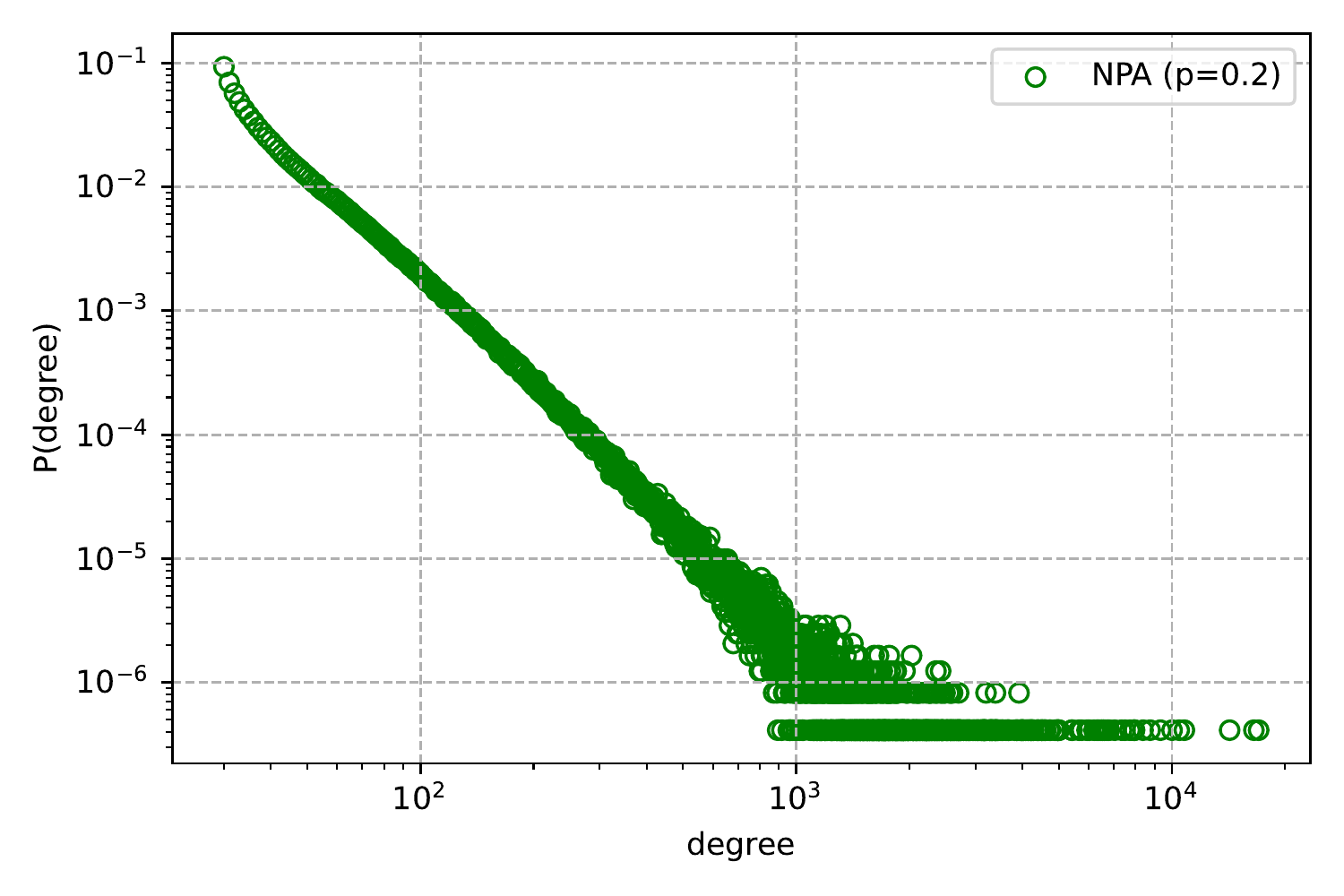}
\caption{Degree distribution in the multi-state road network, using the NPA model with a dropout probability of 0.2 and $m=30$.}
\label{fig:multistate}
\end{figure}

The experiments we have performed so far have been limited to the road networks of individual states. However, Milgram's small-world experiments were performed across multiple states. For this reason, we also performed experiments on the combined road networks of Virginia, Washington, D.C., Maryland, Delaware, New Jersey, New York, Connecticut, and Massachusetts. For $m=30$, we found that the average hop length was $\approx 8.06$, and when we introduced a dropout probability of $p=0.2$, the average hop length was $\approx 7.15$. In \Cref{fig:multistate}, we provide the resulting degree distribution of this road network when the NPA model with a dropout of $p=0.2$ was used.

\subsection{Key Participants}

We also considered the importance of key participants in performing
greedy routing, 
as shown in Figure~\ref{fig:people},
which motivated the NPA model in the first place.

Having a long-tailed
degree distribution could be benefiting the routing phase, as we
know that having more links per vertex improves the asymptotic bound
of Kleinberg's model.  

\begin{figure}[!hbt]
\includegraphics[width=\columnwidth]{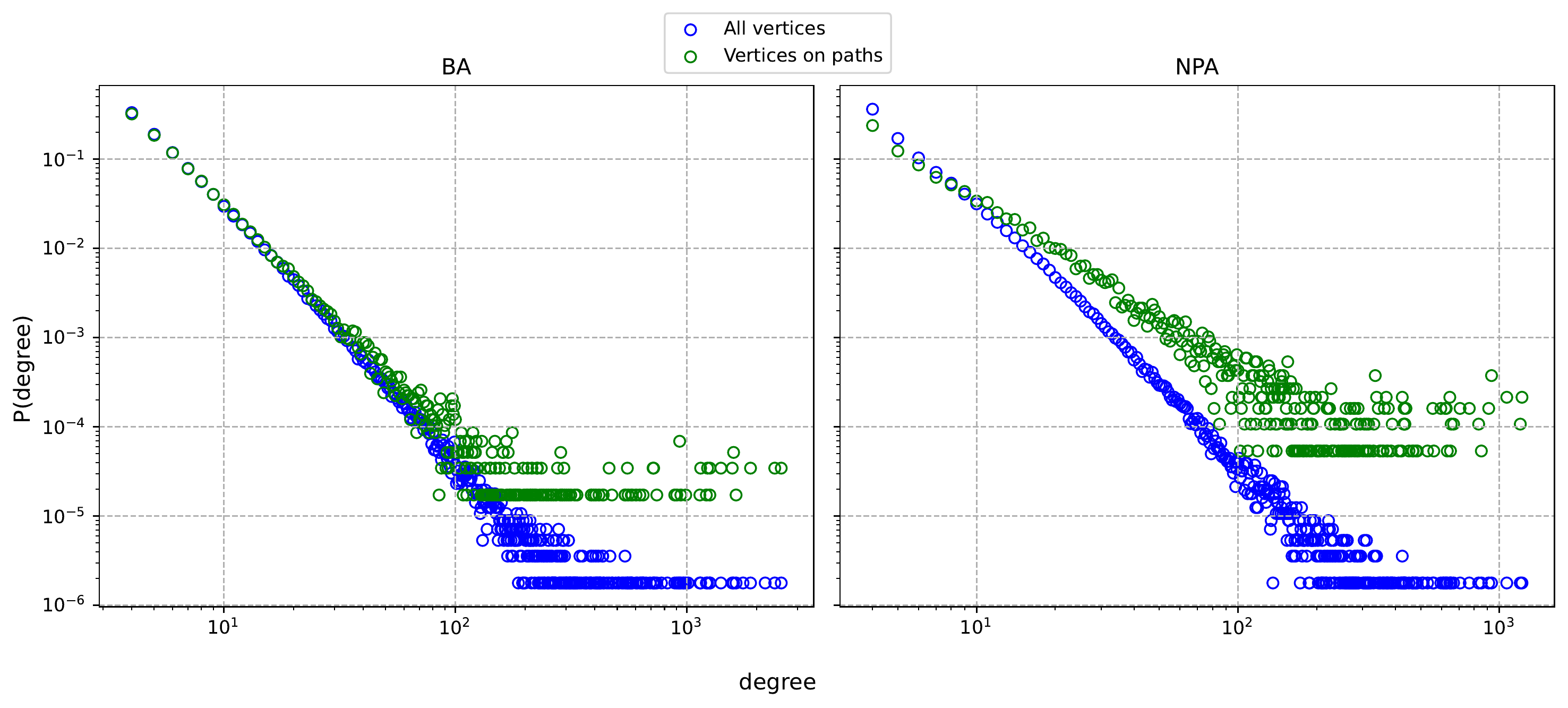}
\vspace*{-12pt}
\caption{Degree distributions in the Washington road network for vertices in the whole network, and vertices visited during the routing phase, using the BA and NPA models with $m=4$.}
\label{fig:path-dd}
\end{figure}

In \Cref{fig:path-dd}, we compare the degree
distribution of vertices that were used during the routing phase
with the degree distribution of the whole network for both the NPA
and BA models. We can see that for the NPA model, high-degree
vertices are being better utilized during an instance of the routing algorithm
compared to the BA model.

\section{Conclusion}
We introduced a new small world model, the Neighborhood Preferential
Attachment model, which combines elements of both Kleinberg's model
and the Barab{\'a}si-Albert model, and experimentally outperforms both
models in terms of the average hop length. 
Importantly, our model is built using real-world distances from nodes
in a road network rather than 
vertices in a square grid or random points on a sphere.

\subsection{Future Work}
For future work, given our experimental results,
it would be interesting to perform
a mathematical analysis of our model, e.g., to see whether our model
has an asymptotic bound on the expected hop length that is 
$o(\log^2 n)$.
Another interesting question is
whether the power law exponent of the degree
distribution differs from the Barab{\'a}si-Albert model in the limit
of the size of the network, or what the diameter of graphs generated
by our model is. 
Yet another interesting problem is whether Kleinberg's
lower bounds for the standard model when the clustering coefficient
is $\neq 2$ still holds for our model.

\section*{Acknowledgements}
This research was supported in part by NSF Grants 1815073 and 2212129.
We also thank David Eppstein for helpful discussions regarding topics
in this paper.

\bibliographystyle{plain}
\bibliography{references}

\end{document}